# Product Cycle and the Industry Cycle

Gerasimos Lianos[1] and Igor Sloev[2]

## Abstract

We consider the problem of dynamic product portfolio choice over a new product (market introduction) and an old product (market stretching) of a firm in a competitive industry. Under complete financial markets, creative destruction is countercyclical. Under incomplete financial markets, creative destruction can be procyclical, (the weight on the old product countercyclical), if the combination of the persistence of industry cycle and degree of financial leverage is high enough. Our model generates an S-curve for creative destruction (measured against shocks in demand), offering alternative foundations for the traditional S-curve of product life cycle.

## PRODUCT CYCLE AND THE INDUSTRY CYCLE

### 1. Introduction and Literature Review

In this letter, we present a theory of the Product Life Cycle (PLC) over the Industry Cycle, with the purpose of integrating business and economic approaches. The (PLC) describes the evolution of products through introduction, growth, maturity, and decline. Although commonly attributed to Levitt (1965), the concept has biological origins. Levitt described the familiar S-shaped sales/profit curve, while Enis and Grace (1977) argued that, unlike biological organisms, products do not follow a predetermined life cycle. Instead, firms actively shape product longevity through strategic decisions regarding introduction, extension, or withdrawal. Since products define industries, the evolution of industry demand is closely linked to product life cycles.

[1] Corresponding author. Higher Colleges of Technology; Abu Dhabi College. Abu Dhabi; UAE. glianos@hct.ac.ae
[2] European University at St Petersburg. St Petersburg. Russia. sia@eu.spb.ru

The PLC literature has evolved from descriptive models (Cox (1967)) toward diffusion theory. Bass (1969) explained product adoption through innovation and imitation effects, later extended by Rogers (2003), Mahajan, Muller et al. (1990), Peres, Muller et al. (2010), and Norton and Bass (1987). Empirical studies (Polli and Cook (1969); Rink and Swan (1979)) found that while the PLC is useful, product trajectories vary considerably. Critics (Dhalla and Yuspeh (1976)) argued that rigid stage models can mislead managers. Day (1981) reinterpreted the PLC as a strategic rather than deterministic framework.

Research on new-product success emphasizes product advantage, market potential, organizational capabilities, and launch timing (Hauser, Tellis et al. (2006); Henard and Szymanski (2001); Steenkamp and Gielens (2003)). Pioneer advantages may arise from learning, distribution, and switching costs, although they are often overstated (Golder and Tellis (1993); Szymanski, Troy et al. (1995); Tellis and Golder (2002)). Innovation research also shows that incumbents often hesitate to cannibalize successful products (Bower and Christensen (1996); Henderson and Clark (1990); Chandy and Tellis (1998)).

Firms frequently prolong product life. Success depends largely on brand fit and marketing support (Aaker and Keller (1990); Keller and Aaker (1992); Reddy, Holak et al. (1994); Völckner and Sattler (2006)). Product elimination and replacement are active portfolio decisions (Avlonitis (1983, 1985)). Portfolio management allocates scarce innovation resources across products (Cooper, Edgett et al. (1999)), while product-line design balances market segmentation against production complexity (Moorthy (1984); Kekre and Srinivasan (1990); Bayus and Putsis (1999); Draganska and Jain (2005)).

Research on marketing over the business cycle shows that fluctuations in aggregate demand affect marketing effectiveness, product strategy, and innovation (Dekimpe and Deleersnyder (2018); Deleersnyder, Dekimpe et al. (2004)). New-product introductions respond strongly to demand conditions (Devinney (1990); Axarloglou (2003); Radas and Shugan (1998)), while launches during recessions may outperform when firms possess sufficient resources (Talay, Pauwels et al. (2024)). Marketing and R&D investments during recessions depend on firm characteristics (Srinivasan, Rangaswamy et al. (2004); Srinivasan, Lilien et al. (2011); Steenkamp and Fang (2011); Mizik (2010)). Downturns also increase private-label demand (Lamey, Deleersnyder et al. (2007, 2012); Dubé, Hitsch et al. (2018)), while the effectiveness of pricing and advertising changes over the business cycle (Van Heerde, Gijsenberg et al. (2013)).

The macroeconomic literature on Schumpeterian growth provides a complementary angle. Schumpeter (1934, 1942) viewed innovation and business cycles as products of creative destruction. Endogenous growth theory formalized this idea (Romer (1990); Grossman and Helpman (1991); Segerstrom, Anant et al. (1990); Aghion and Howitt (1992, 1998, 2009)). The relationship between recessions and innovation remains debated. Recessions may promote restructuring (Caballero and Hammour (1994, 1996); Aghion and Saint-Paul (1998)), yet empirical evidence often finds R&D to be procyclical (Barlevy (2007); Comin and Gertler (2006); Francois and Lloyd-Ellis (2009)). Financial constraints influence innovation choices (Aghion, Angeletos et al. (2010)), and downturns may either cleanse or scar the economy (Ouyang (2009); Foster, Grim et al. (2016); Nanda and Nicholas (2014)).

In this article, we aim to integrate the marketing PLC literature with Schumpeterian growth theory. Our approach builds on Lianos and Sloev (2024), extending earlier marketing models (Lianos and Sloev (2016); Lianos, Sloev et al. (2022)) by linking customer acquisition with new-product introduction and customer retention with product stretching.

## 2. A Model of Product Portfolio Choice

There are $M$ firms in the industry. Each firm lives for two periods. Industry demand is random variable and it is realized at the beginning of second period $a'$, with the law of motion: $\ln(a') = \rho\ln(a) + \varepsilon$, with $0 < \rho < 1$, $\varepsilon \sim N(\mu_\varepsilon, \sigma_\varepsilon^2)$**.** Firms decide over the market stretching and market innovation. Market stretching results in expenditure is $x$ in the old product and innovation in expenditure $z$ in the new product. If the firm $i$ spends $x^i$ in the old product, when the rest of the firms in the industry spend ($x^j$, the firm $i$'s gross realized profits at the end of the first period is: $a\left(x^\theta / \int_0^M \left(x^j\right)^\theta dj\right)$. Similarly, if the firm $i$ spends $z^i$ in the new product, the firm $i$'s gross realized profits at the end of the second period is given by: $E_a(a')\left(z^\eta / \int_0^M \left(z^j\right)^\eta dj\right)$**.** Both expenditures are incurred at the beginning of the first period, after the realization of the random state of industry demand is observed. Investment in the old product gives its return at the end of the first period. Investment in the new product gives its return at the end of the second period.

The representative firm chooses the mix of new and old products to maximize the sum of the two-period cash flows, taking as given the decisions made by the other firms in the industry and the stochastic process of product profitability. The problem of the firm is given by:

$$\max_{x,z}\left\{a\left(x^{\theta}/\int_0^M \left(x^j\right)^{\theta} dj\right) + E_a(a')\left(z^{\eta}/\int_0^M \left(z^j\right)^{\eta} dj\right) - w)\right\}$$

subject to: $x + z = w(a)$, given: $\ln(a') = \rho \ln(a) + \varepsilon$, and $(x^j)$, $(z^j)$, for all $j \in [0, M]$.

The first-order condition with respect to the choices for x and z are given by:

$$a\theta\left(x^{\theta-1}/\int_0^M \left(x^j\right)^{\theta} dj\right) = 1$$

$$E_a(a')\eta\left(z^{\eta-1}/\int_0^M \left(z^j\right)^{\eta} dj\right) = 1$$

The first-order conditions imply:

$$a\theta\left(x^{\theta-1}/\int_0^M \left(x^j\right)^{\theta} dj\right) = E_a(a')\eta\left(z^{\eta-1}/\int_0^M \left(z^j\right)^{\eta} dj\right)$$

In a symmetric equilibrium of the industry, we have the following conditions: $x^j = x,\ z^j = z$. Using these and the condition $E_a(a') = a^{\rho} e^{\mu_\varepsilon + \sigma_\varepsilon^2/2}$ in the first order condition we obtain: $a\theta/Mx = a^{\rho}\eta/Mz$ and the optimal old-to-new technology ratio:

$$\frac{z}{x} = \left(\frac{\eta}{\theta}\right) a^{\rho-1} h$$

where $h = e^{\mu_\varepsilon + \sigma_\varepsilon^2/2}$.

We call the new-to-old technology ratio, $\kappa = z/x$ (extent of) creative destruction. Clearly, $\kappa$ is driven by the shock process and its law of motion can be written as:

$$\ln(\kappa) = (\rho - 1)\ln(a) + \mu_\varepsilon + \frac{\sigma_\varepsilon^2}{2} + c$$

where $c = \ln(\eta/\theta)$.

The quantity $1 - \rho$ is the elasticity of creative destruction with respect to the current industry demand. The elasticity depends on the persistence parameter $\rho$. If the persistence of $a$ is high $(\rho \to 1)$, creative destruction is irresponsive to the state of the demand. If the persistence is low $(\rho \to 0)$, creative destruction moves one percent-to-one percent with the demand.

The optimal demand functions for old and new technologies are given by:

$$x(a) = \left(\frac{1}{1+\kappa}\right) w \; ; \;\; z(a) = \left(\frac{\kappa}{1+\kappa}\right) w$$

Figure 1 shows creative destruction, $\kappa$, as a function of $a$ for $\rho = 0.1$ and $0.9$ ($\theta, \eta = 0.5;\ M = 1, g = 0, \mu_\varepsilon = -\sigma_\varepsilon^2/2$ ).

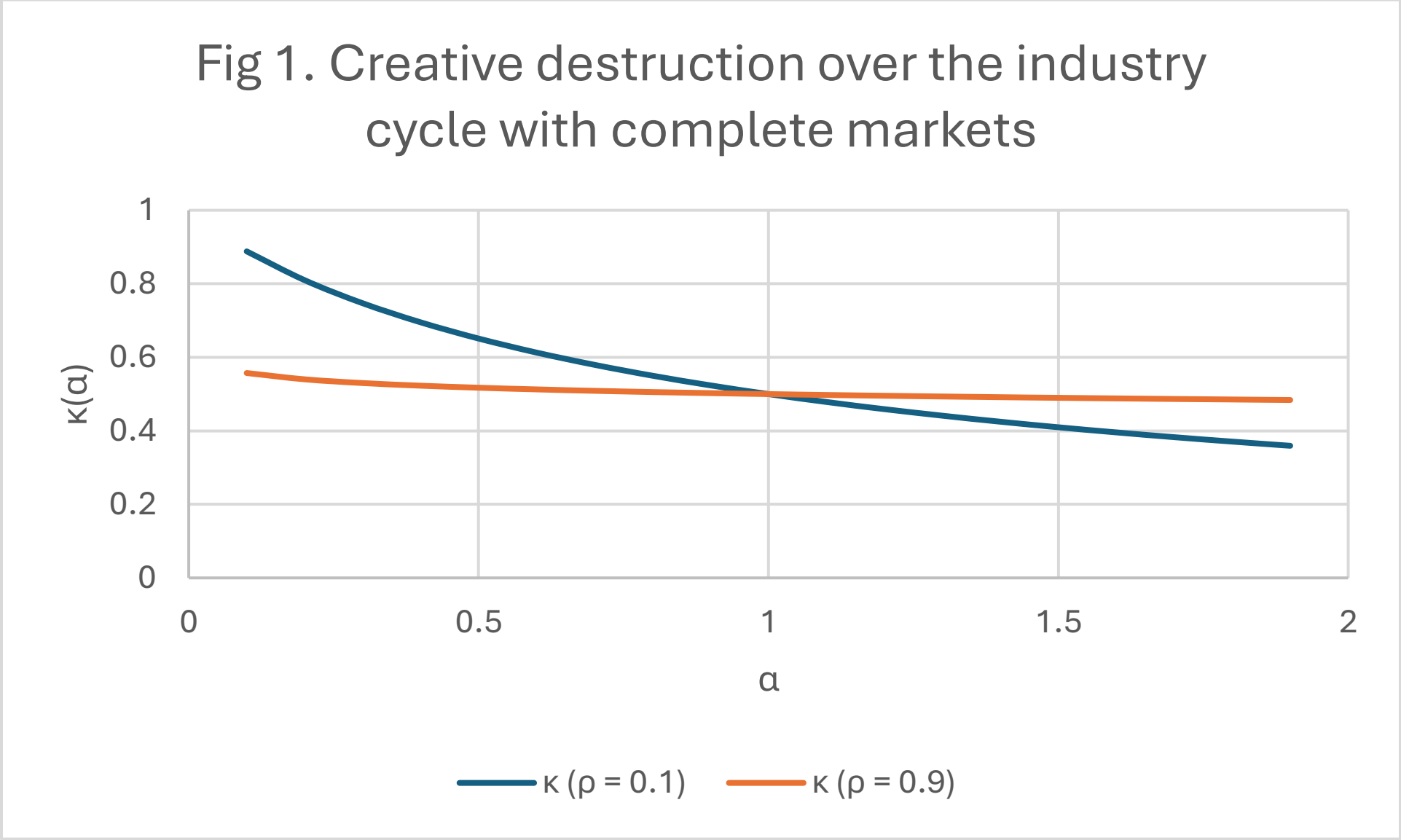

[Figure 1 around here]

### Proposition 1. Creative destruction with complete markets

Creative destruction $\kappa(a)$ decreases with industry demand $a$.

### Proof

(a) It is: $dln(\kappa)/da = (\rho - 1) < 0$, as $\rho < 1$. Hence, $\kappa'(a) < 0$.

Proposition 1 captures the Schumpeterian idea of the "cleansing effect" of economic recessions, stating the recessions drive creative destruction. With small industry demand the opportunity cost of substituting new product for old product is low. Complete markets means the firm does not need to worry about current cash flow.

## 3. Incomplete markets

After the allocation of investment and before the realization of the second period's return, a shock occurs that affects the success of the innovation in the second period. The occurrence of the shock requires additional expenditure for the firm to continue with the new product. The survival of the firm depends on the ability of the firm to finance the additional expenditure, which requires cash flow or borrowing. However, the latter possibility is conditioned by the ability of the firm to leverage existing equity, hence the cash flow generated from the old product.

After the allocation of the total portfolio value, the random liquidity shock $s$ is realized. It is $s \in (0, \infty)$, with distribution function $F(b) \equiv Prob(s \leq b)$. We assume that $lnF(b) \approx \varphi ln(b)$, so that $\varphi$ is the local elasticity of $F$. This means that $F(b) = b^{\varphi}$. After the shock is realized, the firm can continue with the development of the new product only if and only if $s \leq \pi^{\lambda}$, where $\pi$ is the cash flow from the realization of the old product return, [that is, $\pi = a\left(x^{\theta}/\int_0^M (x^j)^{\theta}\, dj\right)$] and $\lambda$ is the financial leverage parameter. It is $\lambda > 0$. The probability of the firm continuing into the second period is: $Prob\left(\zeta \leq \pi^{\lambda}\right) = \pi^{\lambda\varphi}$.

The problem of the representative firm in the industry is to find the allocation of investment into old product $x$ and new product $z$ to maximize the sum of cash flows to the firm. The firm solves the following problem:

$$\max_{x,z}\left\{a\left(x^{\theta}/\int_0^M (x^j)^{\theta}\, dj\right) + E_a(a')\left(z^{\eta}/\int_0^M (z^j)^{\eta}\, dj\right)\left(ax^{\theta}/\int_0^M (x^j)^{\theta}\, dj\right)^{\lambda\varphi} - w\right\}$$

subject to: $x + z = w(a)$, given: $\ln(a') = \rho \ln(a) + \varepsilon$, and $(x^j)$, $(z^j)$, for all $j \in [0, M]$, $\varepsilon \sim N(\mu_{\varepsilon}, \sigma_{\varepsilon}^2)$.

The first-order condition with respect to the choices for x and z are given by:

$$a\theta\left(x^{\theta-1}/\int_0^M \left(x^j\right)^\theta dj\right) + E_a(a')\left(z^\theta/\int_0^M \left(z^j\right)^\theta dj\right)\lambda\varphi\left(ax^\theta/\int_0^M \left(x^j\right)^\theta dj\right)^{\lambda\varphi-1}$$

$$\left(\theta\, ax^{\theta-1}/\int_0^M (x^j)^\theta\, dj\right) = 1$$

$$E_a(a')\eta\left(z^{\eta-1}/\int_0^M \left(z^j\right)^\eta dj\right)\left(ax^\theta/\int_0^M (x^j)^\theta\, dj\right)^{\lambda\varphi} = 1$$

In a symmetric equilibrium of the industry, we have the following conditions: $x^j = x,\ z^j = z$. Using these and the condition $E_a(a') = a^\rho e^{\mu_\varepsilon+\sigma_\varepsilon^2/2}$ in the equations above, we obtain the creative destruction, $\kappa(a) = \left(\frac{z}{x}\right)(a)$:

$$\kappa(a) = Aa^{\rho+\lambda\varphi-1}/\left(1 + Ba^{\rho+\lambda\varphi-1}\right)$$

where: $A = (\frac{\eta}{\theta})e^{\mu_\varepsilon+\sigma_\varepsilon^2/2}$ and $B = \lambda\varphi e^{\mu_\varepsilon+\sigma_\varepsilon^2/2}$.

As before, the investment functions are given by:

$$x(a) = \left(\frac{1}{1+\kappa}\right)w\ ;\ \ z(a) = \left(\frac{\kappa}{1+\kappa}\right)w$$

Figure 2 shows $z/x$, as a function of $a$ for $\rho = 0.1$ and $0.9$ and $\varphi = 0.5$ ($\mu = 0.5;\ \theta = \eta = 0.5;\ M = 1$).

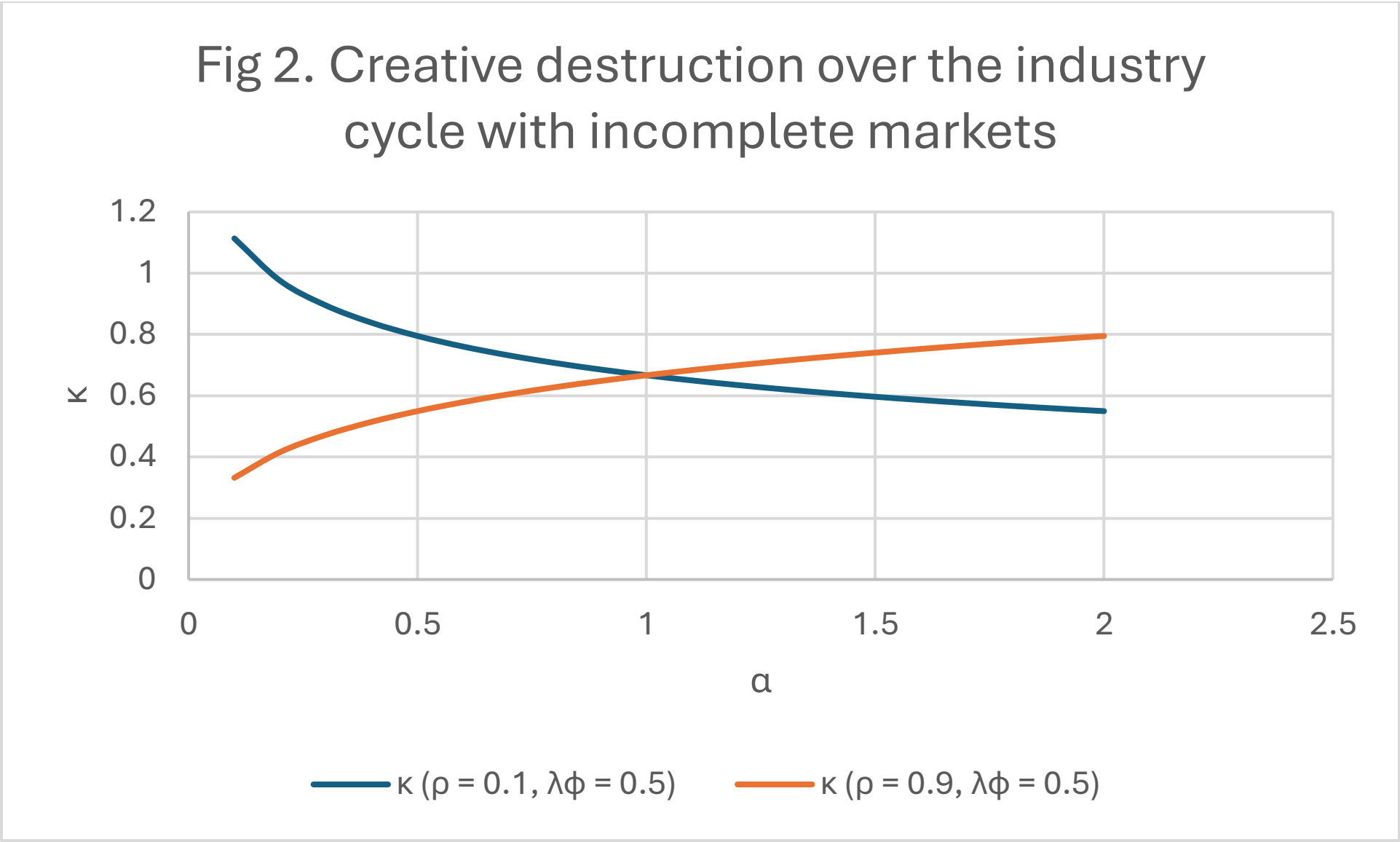

[Figure 2 around here]

**Proposition 2. Creative destruction with incomplete markets**

Creative destruction $\kappa(a)$ increases with industry demand $a$ if and only if $\rho + \lambda\varphi > 1$.

**Proof**

It holds: $\kappa'(\alpha) = (\rho + \lambda\varphi - 1)a^{\rho+\lambda\varphi}A/(a + a^{\rho+\lambda\varphi}B)^2$. Therefore, $sign[\kappa'(\alpha)] = sign(\rho + \lambda\varphi - 1)$.

Proposition 2 states that creative destruction is procyclical, if the combined business cycle persistence and weighted financial leverage are sufficiently high. With incomplete markets, if the leverage capacity of the firm is low, the firm becomes dependent on current cash flow and the new product will be introduced during an industry expansion.

## 4. Endogenous Firm Entry

We consider the possibility that the number of firms in the industry depends on the state of the industry demand**.** The empirical evidence suggests that the number of firms increases in expansions and it decreases in contractions.

We will assume that the number of firms in the first period is $M$ and the number of firms in the second period is $M'$. The expected profit of a firm at the equilibrium of the industry as given by:

$$\frac{a}{M} + \frac{a^{\rho}}{M'}\left(\frac{\alpha}{M}\right)^{\lambda\varphi} - w = 0$$

The function $I(M;a) = \frac{a}{M} + \frac{a^{\rho}}{M'}\left(\frac{\alpha}{M}\right)^{\lambda\varphi} - w$ is continues in $M'$and it satisfies: $I(0) = \infty,\ I(\infty) = -w,\ \partial I(M')/\partial M' < 0.$ This means that there is $M'$that solves the equation $I(M';a) = 0.$ That is, for $M$ and $M'$ number of firms in the industry, firms make zero expected profits. Furthermore, we have: $\frac{\partial I}{\partial M'} = -\left(\frac{\alpha^{\rho}}{M'^2}\right)\left(\frac{\alpha}{M}\right)^{\lambda\varphi} < 0.$ From the Implicit Function Theorem, there is a function $M' = M'(a)$ implicitly defined by $I(M;a) = 0$ and it holds: $\frac{dM'}{da} = -\frac{\partial I}{\partial a}/\frac{\partial I}{\partial M'} = -\left[\frac{1}{M} + \frac{(\rho+\lambda\varphi)\alpha^{\rho+\lambda\varphi-1}}{M'M}\right]/\left[-\left(\frac{\alpha^{\rho}}{M'^2}\right)\left(\frac{\alpha}{M}\right)^{\lambda\varphi}\right] > 0$.

Given the above, we assume that the function $M'(a)$ is defined locally as $M'(a) = Ma^{\gamma}, \gamma > 0.$

With an endogenous number of firms, creative destruction becomes:

$$\kappa(a) = Aa^{\rho+\lambda\varphi-\gamma-1}/\left(M^{\lambda\varphi} + Ba^{\rho+\lambda\varphi-\gamma-1}\right)$$

Where, as before: $A = (\frac{\eta}{\theta})e^{\mu_{\varepsilon}+\sigma_{\varepsilon}^2/2}$ and $B = \lambda\varphi e^{\mu_{\varepsilon}+\sigma_{\varepsilon}^2/2}$. Then Proposition 2 becomes:

## Proposition 3 Creative destruction with firms' entry

$\kappa$ increases with $a$ if and only if $\rho + \lambda\varphi - \gamma > 1$

Proposition 3 expresses the idea that product introduction depends on industrial competition. If $\gamma > 0$ (procyclical entry), for $\kappa$ to be procyclical $\rho + \lambda\varphi$ must be greater than $1 + \gamma$. In this case, during contractions, firms exit and profit shares rise, which makes innovation more profitable. The Schumpeterian “cleansing effect of recessions” takes hold. If $\gamma < 0$ (countercyclical entry), the procylicality condition for creative destruction, $\rho + \lambda\varphi - \gamma > 1$, becomes easier to hold and $\kappa$ tends to become procyclical.

## 5. Demand Shocks and Product Portfolio Choice

Let $u = \mu + \sigma^2/2$. Creative destruction $\kappa(u)$ increases with $u$. Let $m = \rho + \lambda\varphi - \gamma - 1.$ It is: $\kappa(u) = \frac{A}{B}\frac{Be^m e^u}{1+Ba^m e^u} = \frac{A}{B}\frac{1}{1+e^{-[u+\ln(Ba^m)]}}$ . This is a logistic curve in $u$, with inflection point at $u^* = -\ln(Ba^m)$ for which $\kappa(u^*) = \frac{A}{2B}$. The lower horizontal and upper horizontal asymptotes are $\kappa = 0$ and $\kappa = \frac{A}{B}$. We call this curve the S-Curve of Creative Destruction (Product Portfolio Choice).

The model gives rise to an S-curve of creative destruction. Traditional “S-curves” of the product cycle show an S-pattern of revenues of an existing product over time. In the present model, instead of time, the crucial variable is industry demand. Since, we have the new product choice $z = \kappa/(1+\kappa)w$ and the old product choice $x = 1/(1+\kappa)w$, for constant investment budget, the model implies an S-curve for the new product and an inverted S-curve for the old one. However, the inverted S-curve for the old product when old product is measured against industry demand is equivalent to a regular S-curve for the old product when the product is measured against time (under the implicit assumption of product life cycle models that the passage of time bring a waning of the demand for the old product).

Figure 3 shows $\kappa$, as a function of $u$ for $\alpha = 1$ (values of $\rho$ and $\gamma$ are immaterial), $M = 1$, and values of $\lambda\varphi \; = \; 0.5 \quad 1.5$.

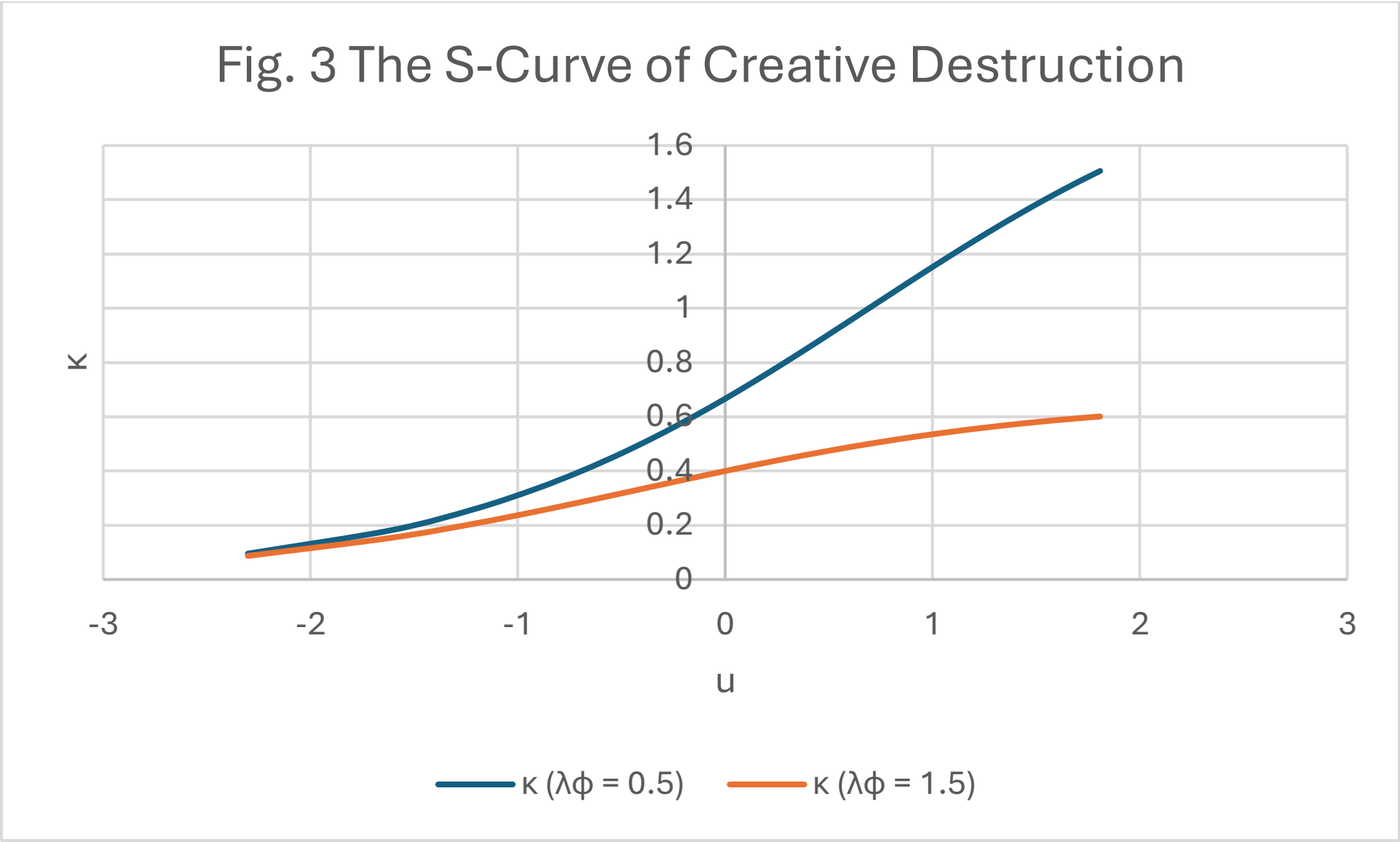

[Figure 3 around here]

## 6. Conclusions

Product life cycles are endogenous rather than predetermined. Product age alone does not determine life-cycle stage, which depends on demand, industrial structure, technological innovation and the industry cycle. In this letter, we have presented a Schumpeterian model of the optimal Product Lifetime Cycle. Our model identifies industry demand as the key driver of product life cycles. Demand influences market size, profitability and financing. During booms, expanding demand encourages innovation, but it increases the profitability of existing products, making firms reluctant to cannibalize them. During recessions, weaker demand discourages risky launches, especially for financially constrained firms, although firms with sufficient financial slack may innovate countercyclically as competition and launch costs decline. Our model distinguishes product stretching from new-product introduction under two different financial environments. With perfect capital markets, booms extend products' lifecycles, while recessions shorten lifecycles. Under imperfect financial markets, constrained firms tend to stretch products' lifecycles during recessions and postpone innovation until expansions. Our model generates an S-curve of creative destruction.